# Macroscopic quantum effects for classical light

N.I. Petrov

Lenina str., 19-39, Istra, Moscow region, Russia 143500

E-mail: petrovni@mail.ru

Macroscopic quantum optical effects (Schrodinger cat states, squeezing, collapse and revival) for light beams propagating in an inhomogeneous linear medium are demonstrated using exact analytical solutions of wave equation. It is shown that the coherent superposition of macroscopically distinguishable states is generated via mode interference from an initial off-axis single wave-packet. Squeezed cat states with a fidelity > 99% arise periodically and disappear rapidly within limited intervals of a propagation distance. Collapse and revival of wave packets at long-term non-paraxial evolution due to mode interference is demonstrated. Oscillations of the beam trajectory with extremely small amplitude of the order of $10^{-19}$ *m* which is typical of the estimated displacement caused by cosmic gravitational waves in gravity-wave detectors occur.

Quantum-optical analogies have been used in the recent years to understand many quantum phenomena at the macroscopic scale. In spite of existence of the nonequivalence of classical and quantum mechanics and optics, very close analogies can be found between classical optics and quantum physics. It is well known that the paraxial beam of light, a purely classical object, obeys the equation formally identical to the Schrodinger equation, except that the evolution follows the axial direction instead of time and the wavelength plays the role of Planck's constant [1]. There are also analogies between classical optics and relativistic theory of particles. Nonparaxial propagation of a scalar optical beam is described by an equation similar to the relativistic equation of Klein-Gordon. Such analogies allow us quantum phenomena by means of light propagation in classical systems to be mimicked. This opens up possibilities to visualize quantum phenomena, such as Schrodinger cat's state, squeezing and collapse and revival, at a macroscopic level. These phenomena are of a great interest in many fields of physics such as quantum optics, quantum computation and precision measurements [2, 3]. It was shown in [4] that Schrodinger's catlike states [5] are generated from an initial coherent state propagating through a Kerr nonlinear medium. The creation and measurement of coherent and Schrödinger-cat states have been performed in [6, 7]. Squeezed light is considered to be used in various fields, e.g., precision measurement like gravitational wave detection [8] and photonic quantum information processing with continuous variables [9]. Squeezed light can be generated in various methods [10-13]. In [14] a method to produce quantum superposition of squeezed coherent states with arbitrary large amplitudes was proposed and experimentally demonstrated. In [15] a gedanken experiment in which the collapse and revival of a coherent state can be observed is realized.

In this paper, the dynamics of squeezing, generation of Schrodinger's catlike states and collapse and revival effects at the nonparaxial propagation of a classical light beam in a linear optical medium are demonstrated. Exact analytical expressions for the moments of observables (trajectory, momentum, and their dispersions) for nonparaxial wave beams propagating in a graded-index medium are obtained using a quantum-theoretical approach of coherent states.

A quantum-theoretical method of coherent states was used in [16] for consideration of a beam propagation in a graded-index medium beyond the paraxial approximation. The first order corrections for the beam trajectory and width were obtained. However, it is necessary to find exact solutions in order to analyze a long-term non-paraxial propagation of wave beams. In [17] the exact analytical expressions for the trajectory and width of nonparaxial wave beams were obtained.

The propagation of radiation in one dimensional media (e.g. planar waveguide) is described by the Helmholtz equation for the monohromatic component of electric field $\vec{E}(x,z)$ following from Maxwell equations:

$$\frac{\partial^2 \vec{E}}{\partial x^2} + \frac{\partial^2 \vec{E}}{\partial z^2} + k^2 n^2(x,z)\vec{E} = 0, \tag{1}$$

where $k = \frac{2\pi}{\lambda}$ is the wavenumber, $n(x,z)$ is the refractive index of medium.

In the case of homogeneous medium in longitudinal direction $z$ the equation (1) may be reduced to the equivalent Shrodinger equation for the reduced field $\psi(x)$:

$$\hat{H}\psi(x) = \varepsilon\psi(x) \tag{2}$$

where $\varepsilon$ and $\psi(x)$ are the eigenvalue and eigenfunction of the Hamiltonian

$$\hat{H} = -\frac{1}{2k^2}\frac{\partial^2}{\partial x^2} + \frac{1}{2}\left(n_0^2 - n^2(x)\right) \tag{3}$$

Evolution of the field $E(x,z)$ is determined by the propagation constant $\beta(\varepsilon)$:

$$E(x,z) = \hat{U}\psi(x), \quad \hat{U} = \exp(i\hat{\beta}z), \quad \beta(\varepsilon) = kn_0\left(1 - \frac{2\varepsilon}{n_0^2}\right)^{1/2}.$$

Here the operator of the propagation constant is introduced

$$\hat{\beta} = kn_0\left(1 - \frac{2\hat{H}}{n_0^2}\right)^{1/2} = kn_0\left(1 - \frac{\hat{H}}{n_0^2} - \frac{\hat{H}^2}{2n_0^4} - \frac{\hat{H}^3}{2n_0^6} - \cdots\right), \tag{4}$$

the eigenvalues of which determine the propagation constants $\beta(\varepsilon)$.

Note, that the eigenvalues spectrum is non-equidistant, which leads to the effective nonlinearity. Thus the solution of the Helmholtz equation (1) in this case may be reduced to the solution of the Heisenberg equation for the operators, the average values of which determine the parameters of the investigated beam, for example, the coordinate and width of the beam.

Consider the propagation of radiation in homogeneous in the longitudinal direction $z$ medium with the parabolic distribution of the refractive index in the transverse direction $x$:

$$n^2 = n_0^2 - \omega^2 x^2, \tag{5}$$

where $n_0$ is the refractive index on the axis, $\omega$ is the gradient parameter of the medium.

As an incident beam we consider the coherent states - the wave packets which are the eigenfunctions of the annihilation operator $\hat{a}$:

$$\hat{a}|\alpha\rangle = \alpha|\alpha\rangle, \tag{6}$$

where the eigenvalues $\alpha = \frac{1}{\sqrt{2}}\left(\sqrt{k\omega}x_0 + i\sqrt{\frac{k}{\omega}}p_0\right)$ determine the initial coordinate $x_0$ of the beam center and the inclination angle $p_0 = n\sin\varphi_0$ of a beam to the axis of the medium.

The coherent states represented by Gaussian wave packets were introduced for the first time by Glauber [18] for the one-dimensional stationary quantum oscillator in the connection with quantum optics problems. Essentially, these states are analogous to the Gauss wave packets in a coordinate representation, which were constructed and studied by Schrodinger [19] for the quantum harmonic oscillator as part of his investigation of the relation between quantum and classical descriptions.

The coherent states have the minimum possible width and diffractional angle divergence at the paraxial propagation in the medium (5). The center of such wave packets is moving along the trajectory of a geometrical ray, i.e. in accordance with the ray optics, and the width of a packet is not changed on the propagation. Besides the coherent states are generating functions for the Gauss-Hermite modes of a medium:

$$|\alpha\rangle = e^{-\frac{|\alpha|^2}{2}} \sum_{m=0}^{\infty} \frac{\alpha^m}{\sqrt{m!}} |m\rangle, \qquad (7)$$

where $|m\rangle = \left(\frac{k\omega}{\pi}\right)^{1/4} \frac{1}{\sqrt{2^m m!}} \exp\left(-\frac{k\omega}{2} x^2\right) H_m(\sqrt{k\omega} x).$

The evolution of the beam trajectory is determined by the calculation of the average value of the operator of coordinate:

$$\bar{x}(z) = \langle \psi_\alpha(z) | \hat{x} | \psi_\alpha(z) \rangle = \langle \psi_\alpha(0) | \hat{x}(z) | \psi_\alpha(0) \rangle, \qquad (8)$$

where $\hat{x}(z) = \hat{U}^+ x \hat{U}$ is found from the solution of the Heisenberg equation $\dot{\hat{x}} = i[\hat{x}, \hat{\beta}]$.

Taking into account the quadratic in $H$ terms in the expansion (4) we have for the trajectory and linear angular momentum of a beam [16, 17]:

$$\langle x \rangle_\alpha = \frac{|\alpha|}{\sqrt{k\omega/2}} \exp\left[|\alpha|^2 (\cos \eta z - 1)\right] \cos\left[\left(\frac{\omega}{n_0} + \eta\right) z + |\alpha|^2 \sin \eta z - \theta\right],$$

$$\langle p \rangle_\alpha = |\alpha|\sqrt{2\omega/k}\exp\left[|\alpha|^2(\cos\eta z - 1)\right]\sin\left[\left(\frac{\omega}{n_0} + \eta\right)z + |\alpha|^2\sin\eta z - \theta\right], \qquad (9)$$

where $\eta = \dfrac{\omega^2}{kn_0^3}$, $\alpha = |\alpha|e^{i\theta}$.

It is seen that in contrast to paraxial optics the usual relation between the linear angular momentum and velocity does not exist, i.e. $\bar{p}_\alpha \neq \dfrac{d\bar{x}_\alpha}{dz}$. Note, that the similar inequality between the linear angular momentum and velocity takes place in relativistic quantum mechanics. It is known that the quantum solutions are nonanalytic functions of $\hbar$ at $\hbar = 0$. Here we can see that there is no such singularity at the classical limit $\lambda \to 0$. At $\lambda \to 0$ we have the paraxial solution for $\bar{x}$ and $\bar{p}$.

Exact expressions for the coordinate of the beam center and momentum which take into account all terms in the expansion (4) have the form:

$$\langle x \rangle_\alpha = \frac{|\alpha|}{\sqrt{k\omega/2}}\exp(-|\alpha|^2)\sum_{m=0}^{\infty}\frac{|\alpha|^{2m}}{m!}\cos\left[kn_0\left(\sqrt{1-\frac{2\omega}{kn_0^2}\left(m+\frac{1}{2}\right)} - \sqrt{1-\frac{2\omega}{kn_0^2}\left(m+\frac{3}{2}\right)}\right)z - \theta\right]$$

$$\langle p \rangle_\alpha = \sqrt{\frac{2\omega}{k}}|\alpha|\exp(-|\alpha|^2)\sum_{m=0}^{\infty}\frac{|\alpha|^{2m}}{m!}\sin\left[kn_0\left(\sqrt{1-\frac{2\omega}{kn_0^2}\left(m+\frac{1}{2}\right)} - \sqrt{1-\frac{2\omega}{kn_0^2}\left(m+\frac{3}{2}\right)}\right)z - \theta\right] \quad (10)$$

Notice that by contrast to paraxial approximation the expression for the nonparaxial ray trajectory depends on a wavelength.

The dispersion of the coordinate $\sigma_x$, momentum $\sigma_p$, and their correlation $\sigma_{xp}$ are given by

$$\sigma_x^2 = \Delta x_\alpha^2 = \langle x^2 \rangle - \langle x \rangle^2, \quad \sigma_p^2 = \Delta p_\alpha^2 = \langle p^2 \rangle - \langle p \rangle^2,$$

$$\sigma_{xp} = \frac{1}{2}\langle xp + px \rangle - \langle x \rangle\langle p \rangle. \qquad (11)$$

The evolution of the width of a wave packet is determined by the expression:

$$\Delta x_\alpha^2 = \frac{1}{2k\omega}\left\{1 + 2|\alpha|^2 + 2|\alpha|^2 e^{-|\alpha|^2}\sum_{m=0}\frac{|\alpha|^{2m}}{m!}\cos\left[kn_0\left(\sqrt{1-\frac{2\omega}{kn_0^2}\left(m+\frac{1}{2}\right)} - \sqrt{1-\frac{2\omega}{kn_0^2}\left(m+\frac{5}{2}\right)}\right)z - 2\theta\right]\right\} - \langle x\rangle_\alpha^2,$$

(12)

Analogically, the expressions for the momentum dispersion $\sigma_p$ and correlation between coordinate and momentum $\sigma_{xp}$ can be obtained.

Let us compare the explicit solutions (10) for the expectation values of $\langle x\rangle$ and $\langle p\rangle$ with the solutions presented in (9). Figs 1(a)-(d) show the evolution of $x(z)$ and $p(z)$ as a function of distance for different values of $|\alpha|^2$. Here and below the medium (5) with the gradient parameter $\omega = 7\cdot 10^{-3}$ μm and refractive index $n_0 = 1.5$ is considered. These parameters are reasonable for conventional optical waveguides. For small values of $|\alpha|^2$ the trajectories oscillate with the period $L_{cl} = 2\pi n_0/\omega$ and their amplitudes remain almost constant. For medium and large values of $|\alpha|^2$ the amplitude of the trajectory relaxes toward the waveguide axis and essentially remain static at the value close to zero. It is evident that the long-term periodic revivals of the initial trajectory occur with a period close to $L_{rev} \approx 2\pi/\eta$. Note that $L_{cl} \ll L_{rev}$. Oscillations of the trajectory with the extremely small amplitude of the order of $10^{-19}$ m occur during the evolution of the wave-packet (Fig.1d).

Differences between the solutions (9) and (10) become significant at large $|\alpha|^2$ and long distances (Fig.2). Moreover, if the increase of $|\alpha|^2$ in (9) leads to the sharp decrease of the minimum value $x_{min}$, then in case of (10) the minimum value does not decrease continually. It is followed from (9) that for $|\alpha| = x_0\sqrt{k\omega/2} = 5.6$ the value $x_{min} = 5\cdot 10^{-27}$ m, and $x_{min} = 10^{-19}$ m in case of the exact solution (10) (Fig.2f). The longer the distance, the bigger the difference. This indicates that at a long-term nonparaxial evolution all terms in the expansion (7) should be taken into account. The corresponding phase-space plots (Figs. 2(a),(c),(e)) show that for small values of $|\alpha|^2$ the phase trajectories are close to the circles, and become more complicated for medium and large values of $|\alpha|^2$. The successive revivals occur but the trajectory never returns back to its preceding positions. Essentially, although the trajectory relaxes toward the waveguide axis, it never riches the zero point in a phase space. This indicates that the oscillations of the trajectory with a non-zero amplitude take place. These oscillations are responsible for the revival effect. The minimum values of the amplitudes are decreased with the increasing $|\alpha|^2$. The upper limit of the values of $|\alpha|^2$ is determined by the waveguide parameters, i.e. by its refractive index and transverse dimension.

Figs. 3(a)-(d) show the evolution with the distance of the second moments $\sigma_x$ and $\sigma_p$, and the uncertainty products. These products acquire their initial minimum values at every revival growing to higher values in between revivals. Near the fractional revivals the uncertainty products fall to smaller values. During the revivals, the uncertainty relations of Heisenberg, $\Delta x^2 \Delta p^2 \geq \frac{1}{4k^2}$, and Schrodinger-Robertson [20], $\sigma_x^2 \sigma_p^2 - \sigma_{xp}^2 \equiv \sigma_x^2 \sigma_p^2 (1-r^2) \geq \frac{1}{4k^2}$, where $r = \frac{\sigma_{xp}}{\sigma_x \sigma_p}$ is the correlation coefficient, become an equality. Fig.3c shows the uncertainty product $\Delta x \Delta p$ variation with a distance. This product returns to its initial value at every revival increasing to higher value $(1/2 + |\alpha|^2)/k$ in between revivals. During the fractional revivals, the Schrodinger-Robertson uncertainty product drops to smaller values than the Heisenberg uncertainty product (Fig.3d). In Fig. 4a,b the variations of the correlation between the coordinate and momentum and the correlation coefficient as a function of distance

are presented. The striking feature of collapse and revival effects is the generation of squeezed states at $L = L_{rev}/2$ and $L = L_{rev}$, the squeezing is stronger at $L = L_{rev}/2$ (Fig. 4c). Usually the generation of squeezed states using the parametric excitation in a Kerr nonlinear medium is considered. Here the dynamical squeezing of light during a nonparaxial evolution in an inhomogeneous linear medium via the mode interference occurs. In Fig. 4c the coefficient of the relative squeezing $\gamma = \omega \sigma_x / \sigma_p$ as a function of a propagation distance is presented. The squeezing $\gamma > 7$ occurs at the midways between revivals for initial value $x_0 = 20$ μm.

The survival of the initial state can be examined calculating the autocorrelation function $P(z) = |\langle \psi(x,0) | \psi(x,z) \rangle|^2$, which measures the correlation of the wave function at distance $z$ with its initial state at $z = 0$. Fig. 5 displays the correlation degrees as a function of a distance. The initial peak intensity decreases rapidly, and the fractional and full revivals appear at the determined distances. Note that the full revivals and half order revivals of the correlation degree coincide with the corresponding revivals for the trajectory and width of a beam. The evolution of shapes of the wave packets are shown in Fig. 6. It is seen that the wave packet displays quantum revivals and a coherent state is converted into the coherent superposition of two macroscopically distinguishable states at the midways between revivals. Note that these Schrodinger cat states are appeared at distances where the squeezing takes the highest value. The squeezing, around 9 dB, is achievable for a cat state with $|\alpha| = 5.6$. The fidelity $F = |\langle \psi_{id} | \psi(x,z) \rangle|^2$ between the state $|\psi(x,z)\rangle$ created during the evolution and an ideal Yurke-Stoler cat state [4] is $F > 99\%$ (Fig.7). Cat states are destroyed rapidly owing to interference effects. There are narrow distance intervals where the fidelity periodically takes the values close to 100%. Within the greater part of these intervals the fidelity is close to zero (Fig.7b). Long-term revivals of the fidelity consist of periodical short-term spikes. This indicates that the cat state arises and disappears in a short gap of the distance $\delta z$ which is less than $L_{cl} = 2\pi n_0 / \omega$.

Effects considered may be useful in laser gravitational-wave experiments. It is known that the squeezed light allows high-resolution measurements to be carried out, beating the standard quantum limit [2]. Indeed, using the squeezed states LIGO detector demonstrated the best broadband sensitivity to gravitational waves ever achieved [21]. Note that the multimode waveguide interferometers based on collapse and revival effects of light were demonstrated in [22, 23]. Here we consider the possibility to measure displacements of the order of $10^{-18}$ m which is sufficient for the observation of gravitational waves. To detect such a small mechanical displacement the optical lever consisting of two butt-jointed waveguides (Fig. 8) can be considered. Axis displacement between two waveguides caused by the gravitational wave will affect output parameters of a propagating beam. As an output parameter the phase change or the oscillations and revivals of the transverse momentum of light can be selected. Optical phase shift $\delta\phi = \phi(\delta x) - \phi(0)$ caused by the axis displacement $\delta x$ is determined from the calculation of $\phi(x', z) = \arctan[\operatorname{Im}(\psi(x', z))/\operatorname{Re}(\psi(x', z))]$, where $x' = x - \delta x$.

The wave function of a output beam is determined by

$$\psi(x', z_0 + L) = \hat{U}(x')\psi(x, z_0) = e^{-|\alpha|^2/2} \sum_{m=0}^{\infty} \left( \frac{\alpha^m e^{i\beta_m z_0}}{\sqrt{m!}} \sum_{n=0}^{\infty} a_{mn} e^{i\beta_n L} |n\rangle \right), \quad (13)$$

where $a_{mn} = \int dx' \psi(x'+\delta x)\psi(x')$ are the mode coupling coefficients, $z_0$ is the length of the first waveguide, $L$ is the length of the second waveguide.

The overlap integrals are calculated analytically. Calculations show that the phase shifts of the order $\delta\phi \propto 10^{-6}$ take place, which can be easily observed from measurements. Note it is followed from LIGO interferometer optical parameters that the measurement of $10^{-18}$ *m rms* requires a phase shift measurement of $10^{-9}$ rad *rms* [24, 25]. The sensitivity to the axial displacement is higher when the total length of two waveguides equals to the distance of the Schrodinger cat states formation. This distance also corresponds to the maximum squeezing of the wave packet. Note that the distance at which the squeezing occurs can be decreased significantly if the waveguide with higher value of the gradient parameter is used. The angular momentum change can also be detected as a function of initial axis displacement of a waveguide. The axis displacement of the order of $10^{-18}$ *m* causes the change in the propagation direction of the outgoing beam of $10^{-9}$ *rad*. Such changes in propagation direction can be observed experimentally, which means that the optical lever considered can provide sensitivity necessary to detect gravitational waves from astrophysical sources. Indeed, in [26] a small optical lever is described which can detect a change of $10^{-10}$ *rad* in orientation of a 2 $mm^2$ mirror.

Thus, macroscopic quantum optical effects (collapse and revival, Schrodinger cat states, squeezing) for light beams propagating in an inhomogeneous medium are demonstrated. Usually these effects are considered in nonlinear medium, here these effects are shown in a linear optical medium. Possibility of measurement of extremely small displacements is shown, this can be used in laser gravitational-wave experiments and in ultra-high resolution spectroscopy. Although the results obtained correspond to planar optical waveguides, the effects may also manifest itself in two dimensional waveguides (optical fibers). However, the analysis of these effects in this case is more complicated and will include the solution of Maxwell equations taking into account the polarization effects [27].

In conclusion, classical beam propagation in optical linear medium can simulate quantum phenomena such as squeezing, Schrodinger cat states, collapse and revival. The generation of optical squeezed Schrodinger cat states, which are macroscopically distinguishable, from a single light beam is demonstrated. Oscillations with the extremely small amplitude of the order of $10^{-19}$ m (less than one-thousandth of the radius of a proton ) occur. This value is typical of the displacements caused by cosmic gravitational waves in gravity-wave detectors. Results may be useful in various fields, e.g. in gravitational wave experiments, high precision spectroscopy, and quantum information processing and metrology.

**Figure captions:**

**Fig. 1** Trajectory (solid line) and linear angular momentum (dashed line) of the beam of wavelength $\lambda = 0.63\,\mu m$ as function of distance for $p_0 = 0$ and $x_0 = 15\,\mu m$ (a) and $x_0 = 30\,\mu m$ (c), and their higher resolution plots indicating oscillations with period of $L_{cl}$ (b, d).

**Fig. 2** Phase-space plots for different values of $x_0$: $x_0=1\,\mu m$ (a, b), $x_0=20\,\mu m$ (c, d), $x_0=30\,\mu m$ (e, f). Solid lines correspond to exact solution (10), dashed lines – to (9).

**Fig. 3** The evolution of second order moment $\sigma_x$ (a) and the uncertainty products of Heisenberg (b) and Schrodinger-Robertson (c, d) for $x_0=20\,\mu m$.

**Fig. 4** Position-momentum covariance (a), correlation radius (b), and relative squeezing (c) variations with distance for $x_0=20\,\mu m$. A higher resolution plot of relative squeezing indicating regular spikes with period of $L_{cl}/2$ (d).

**Fig. 5** The autocorrelation function $P(z) = |\langle \psi(x,0)|\psi(x,z)\rangle|^2$ for a value of $|\alpha| = 3.74$.

**Fig. 6** Wave packet intensity distribution evolution with distance for $x_0=20\,\mu m$. Generation of cat state at $L=L_{rev}/2$ (d, j) and revival of initial wave packet at $L=L_{rev}$ (h, k). Wave packet shapes (c, e, g) at fractional revivals.

**Fig. 7** Variation of the fidelity of wave packet with Yurke-Stoler state with distance for $x_0=20\,\mu m$ (a). Higher resolution plot of the fidelity indicating regular spikes of the fidelity with distance $\delta z \ll L_{cl}$ (b).

**Fig. 8** Scheme for coupling a mechanical oscillator's position to optical butt-jointed waveguides $W1$ and $W2$: $L$ – laser, $P$ – photo-detector.

Fig. 1

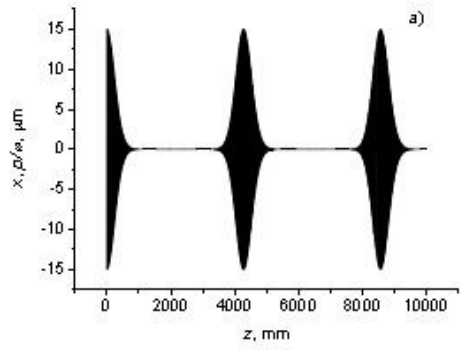 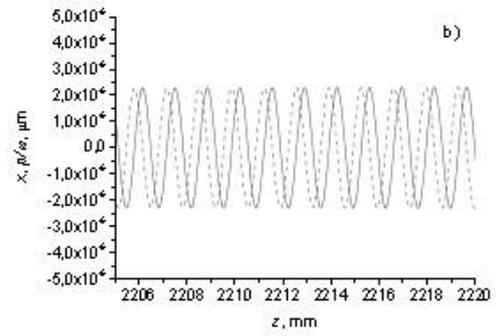

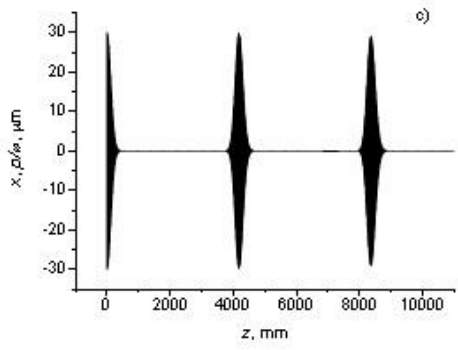 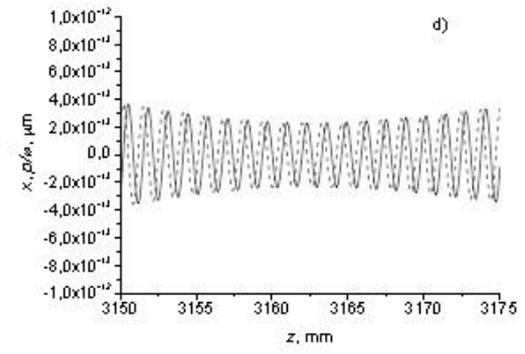

Fig.2

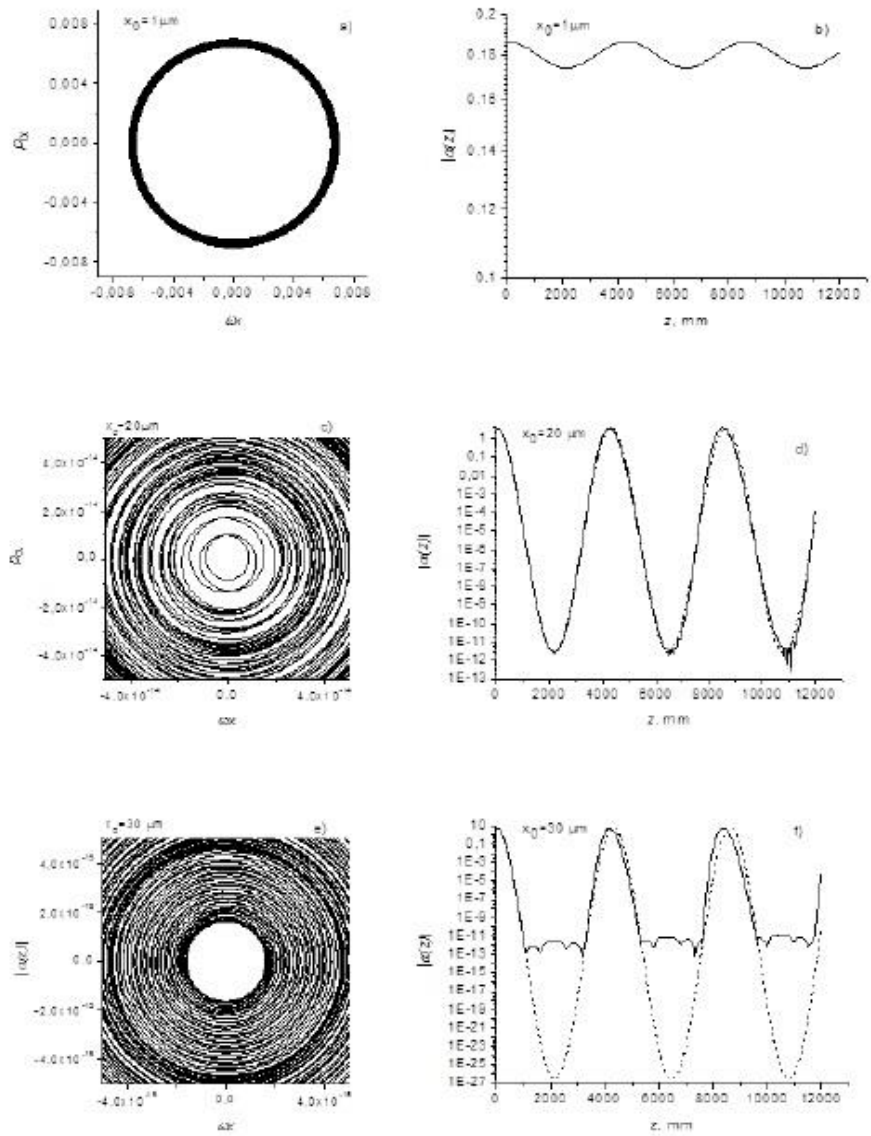

Fig.3

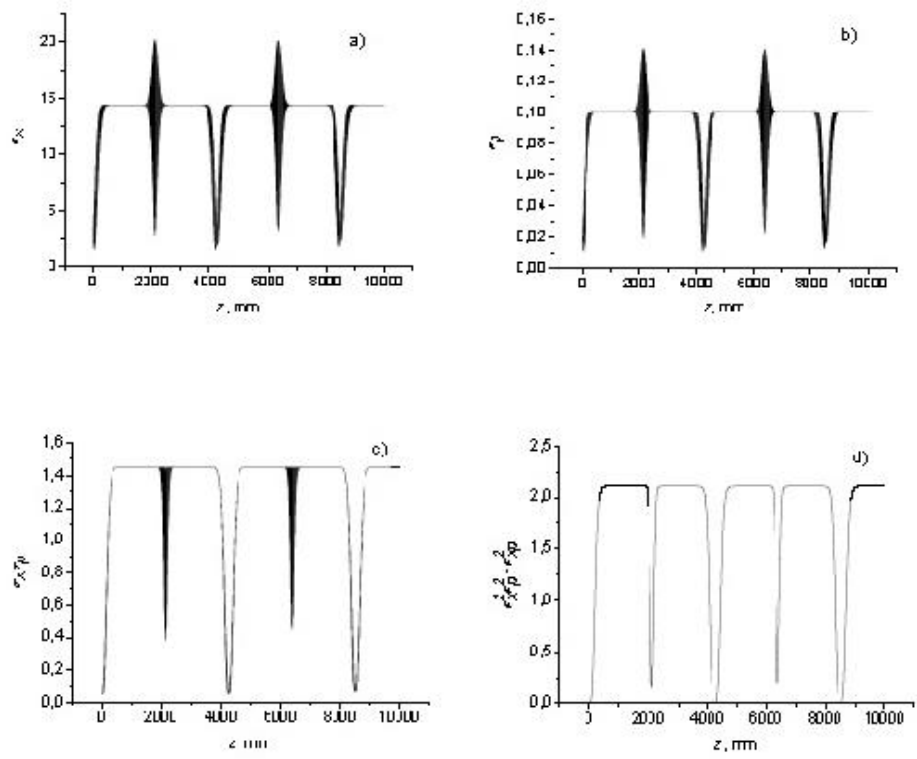

**Fig. 4**

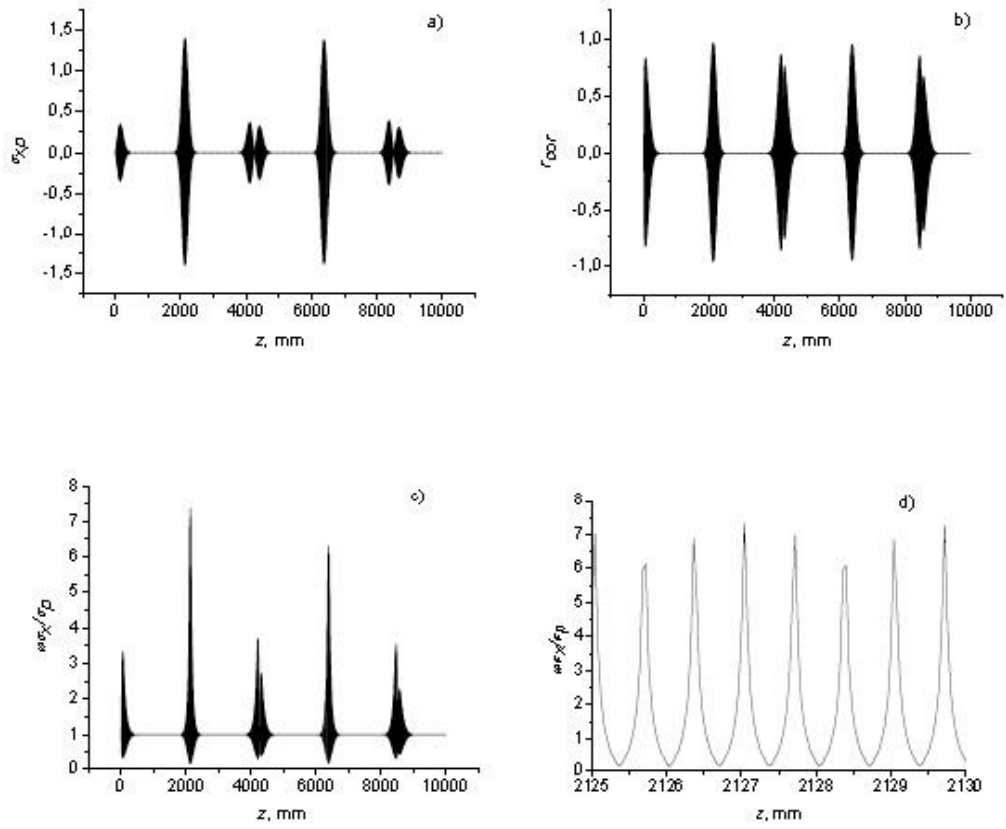

**Fig. 5**

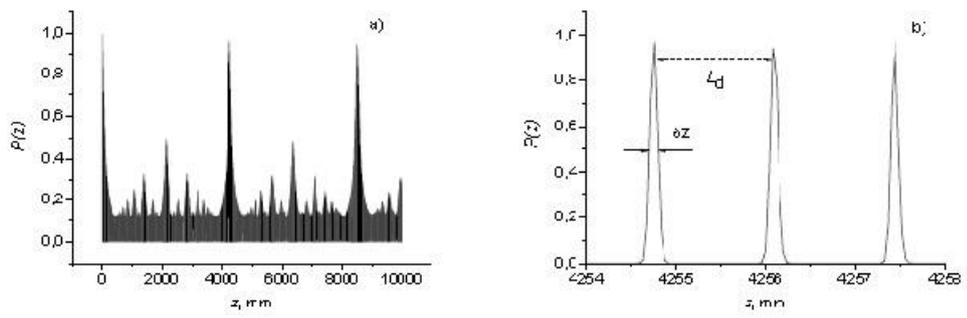

Fig. 6

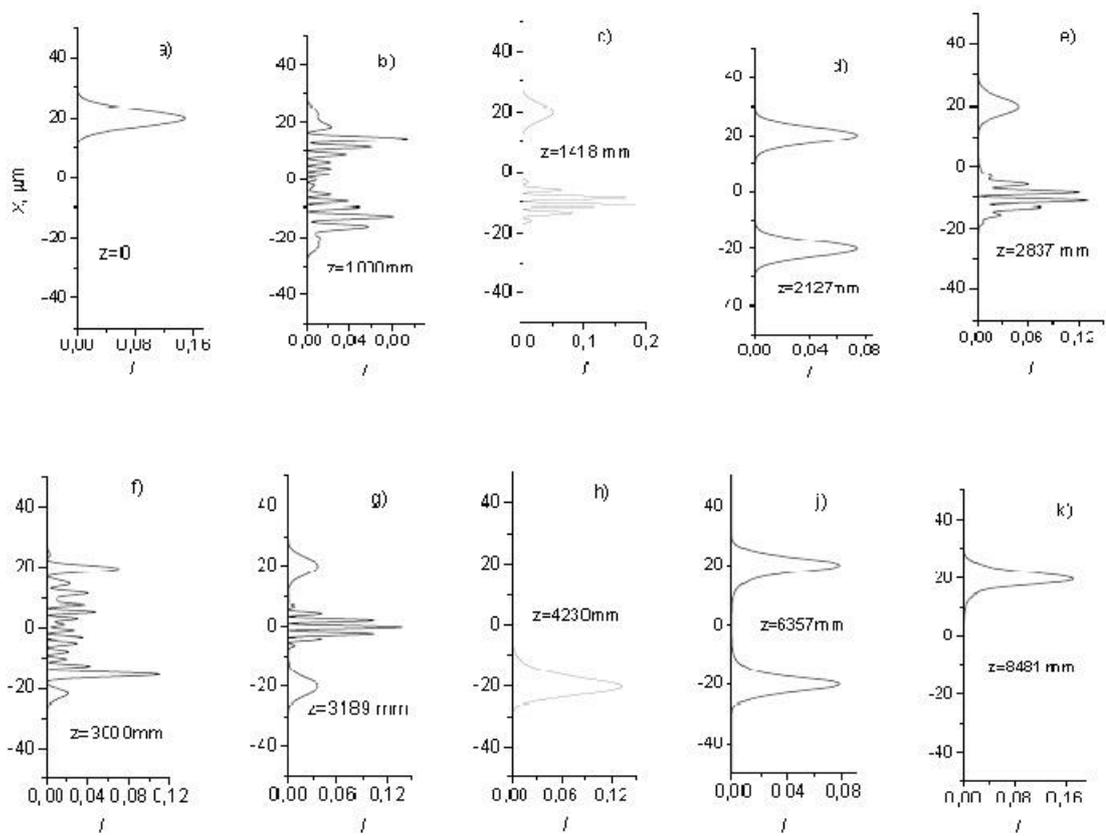

Fig. 7

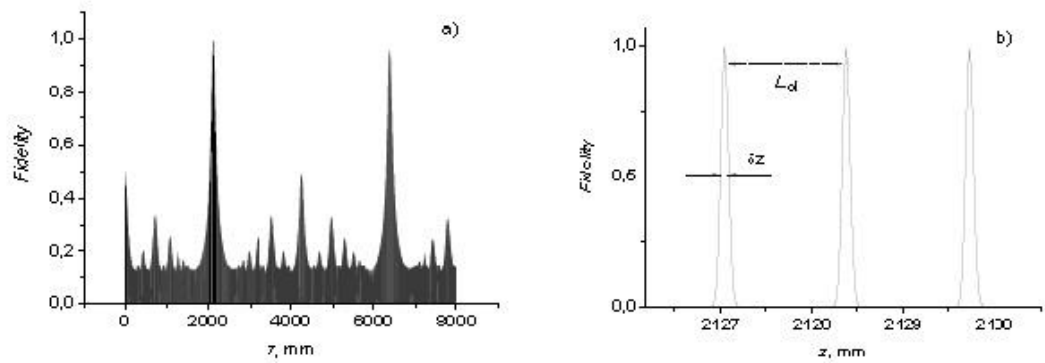

Fig. 8

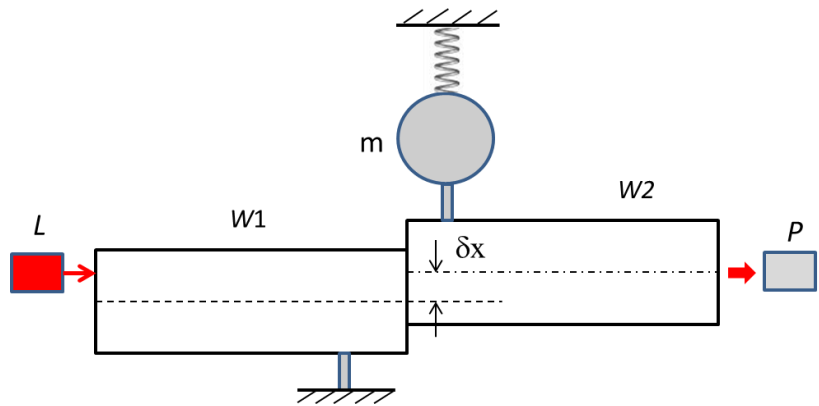